\documentstyle[11pt]{article}
\def\a{\alpha}
\def\b{\beta}
\def\rar{\rightarrow}

\def\le{\left(}
\def\ri{\right)}
\def\t{\theta}
\def\bt{{\bar{\theta}}}
\def\ve{\varepsilon}

\def\f12{\frac{1}{2}}
\def\fra1g2{\frac{1}{g^2}}
\def\Tr{{\rm Tr}}

\def\dis{\displaystyle}
\def\del{\delta}
\def\G{\Gamma}

\def\no{\nonumber}

\def\pd{\partial}

\begin{document}
\begin{titlepage}
\flushright{USM-TH-144}\\
\vskip 2cm
\begin{center}
{\Large \bf {An approach to solve Slavnov-Taylor identity in D4 ${\cal N}=1$ 
supergravity}}
\vskip 1cm  
Igor Kondrashuk\\
\vskip 5mm  
{\it Departamento de F\'\i sica, Universidad T\'ecnica 
Federico Santa Mar\'\i a, \\
Avenida Espa\~{n}a 1680, Casilla 110-V, Valpara\'\i so, Chile} \\
\vskip 5mm 
e-mail:~igor.kondrashuk@usm.cl \\
\end{center}
\vskip 20mm
\begin{abstract}
We consider a particular solution to Slavnov-Taylor identity in 
four-dimensional supergravity. The consideration is performed for 
pure supergravity, no matter superfields are included. The solution 
is obtained by inserting dressing 
functions into ghost part of the classical action for supergravity.
As a consequence, physical part of the effective action 
is local invariant with respect to diffeomorphism and structure groups
of transformation for dressed effective superfields of vielbein and 
spin connection.
\end{abstract}
\end{titlepage}  

\noindent
The classical action of D4 supergravity is of the form 
\begin{eqnarray*}
S_{\rm cl} = -\frac{6}{\kappa^2}\dis{\int~d^4y~d^2{\Theta}~{\cal{E}}R,}
\end{eqnarray*}
where ${\cal{E}}$ is chiral superdensity. We use the notation of Ref. 
\cite{Wess}. Originally in the effective action there are two local 
symmetries. First one is related to diffeomorphism group and 
second symmetry is related to the structure group of rotations in the 
tangent space. In case of present consideration it is $SL(2,C)$ group. 
In general, both the chiral 
density and the chiral superfield $R$ are functions of vielbein ${E_M}^A$ 
and spin connection ${\phi_{MA}}^B$, 
\begin{eqnarray}
S_{\rm cl} = -\frac{6}{\kappa^2}
\dis{\int~d^4y~d^2{\Theta}~{\cal{E}}({E_M}^A,{\phi_{MA}}^B)
R({E_M}^A,{\phi_{MA}}^B)} \label{classical-action-supergravity}.
\end{eqnarray} 
Usually, the chiral superfield $R$ is taken as an independent superfield 
at the classical level. However here we keep it as a function of the 
spin connection and the vielbein that will be variables of integration
in the path integral. The relation between differential 2-form of the 
torsion and 1-form of vielbein is
\begin{eqnarray}
T = {\cal{D}}E,      \label{constraint}
\end{eqnarray}
where ${\cal{D}}$ is a covariant derivative ${\cal{D}} = d + \phi$ with 
respect to structure group
with spin connection inside. This relation will be kept in the path integral
by using a Lagrange multiplier $\pi^\omega,$ that is 
\begin{eqnarray*}
\int~d\pi^\omega~\exp{\int~d^8z~i~E~\pi^\omega{\cal C}^\omega},
\end{eqnarray*}
where ${\cal C}^\omega$ is the constraint (\ref{constraint}), and $\omega$
are indices of representation in the constraint (\ref{constraint}).
The coordinate $z$ is a general coordinate of the supermanifold 
$z^M = (x^M,\t^{\mu},\bt^{\dot{\mu}}).$ The Grassmannian coordinate of the 
manifold does not coincide with the coordinate in the chiral measure. 
The coordinate of the chiral measure $\Theta$ is a function of the 
manifold coordinates and of the vielbein. The torsion $T$ satisfies 
constraints in the tangent space to have flat supersymmetry as a limiting
case in which curvature is absent.  
  
Total action including gauge fixing, FP ghost action, Lagrange 
multiplier at the classical level can be written as      
\begin{eqnarray}
& \dis{S = S_{\rm cl} + S_{\rm gf} + S_{\rm gh} 
  + \int~d^8z~ E~\pi^\omega{\cal C}^\omega} \no\\ 
& = \dis{-\frac{6}{\kappa^2}\int~d^4y~d^2{\Theta}~{\cal{E}}({E_M}^A,{\phi_{MA}}^B)
R({E_M}^A,{\phi_{MA}}^B) - \int~d^8z ~\frac{1}{2\a}\le \pd_M {E_A}^M \ri^2}
  \label{total-action-gravity} \\
& \dis{-\int~d^8z~\Tr\le\frac{1}{\b}\left[\pd_M{\phi^M}(z)\right]^2 \ri - 
\int~d^8z ~i~b^A~\pd_M~{\cal L}_c {E_A}^M } \no\\ 
& \dis{-\int~d^8z~2~\Tr~\le~i~b'(z)\pd_M~{\cal D}^M~c'(z)\ri
+ \int~d^8z~E~\pi^\omega{\cal C}^\omega}. \no
\end{eqnarray}
Here $E = \det{{E_M}^A}$ and ${\cal L}_c$ is Lie derivative of the 
vielbein. It acts on any superfield with world index as  
\begin{eqnarray*}
{\cal L}_c {E_A}^M = c^L\pd_L{E_A}^M - (\pd_L~c^M){E_A}^L,
\end{eqnarray*} 
$c^L$ here is the ghost superfield, $b^A$ is antighost superfield.  
Such a choice of the gauge fixing and the ghost terms means that we fix
the gauge arbitrariness by imposing the condition
\begin{eqnarray*}
{\pd}_M {E_A}^M = F_A, ~~~ \pd_M {\phi^M}_A{}^B = {f_A}^B(z), 
\end{eqnarray*}
where  $F_A$, ${f_A}^B(z)$ are some functions. The first gauge fixing 
condition is to fix the gauge freedom in diffeomorphism group while 
second one is to fix structure group freedom.

The second gauge fixing term and the second ghost term can be made  invariant 
with respect to the diffeomorphism group by construction. Indeed, the gauge 
fixing term can be made invariant by amounting the gauge fixing parameter 
$\beta$ to superfield with property of the density $E$ under diffeomorphism
transformations. The same property can be required for antighost $b'.$   
The first gauge fixing term and the first ghost term in the action  
(\ref{total-action-gravity}) are invariant with respect to 
structure group since the covariant derivative of the vielbein is
\begin{eqnarray*}
{\cal D}_M {E_A}^M = \le\pd_M\del_A^B + {\phi_{MA}}^B\ri~{E_B}^M = 
\pd_M~{E_A}^M +  {\phi_{BA}}^B.   
\end{eqnarray*}
The l.h.s. of this equation and the second term on the r.h.s. are 
covariant with respect to structure group, hence the first term on the 
r.h.s. is also covariant. The first ghost term is covariant with respect
to structure group since the Lie derivative can be written in a covariant
way with respect to structure group form,   
\begin{eqnarray*}
{\cal L}_c {E_A}^M =  {\cal D}^M~c_A + {L_A}^B{E_B}^M,
\end{eqnarray*}
where ${L_A}^B$ is a matrix that takes values in the algebra of 
Lorentz group. The reason for the covariance is the same as in the 
example above. To make use of the diffeomorphism symmetry, we define 
the path integral extended by the dependence on the following external 
sources 
\begin{eqnarray}
& \dis{Z[I,J,~\eta,~\rho,~K,~K'',~L] 
= \int~d{\phi_{MA}}^B~d{E_M}^A~dc^L~db^A~d\pi^\omega~db'~dc'~
 \exp i}\left[\dis{S} \right.  \no \\
& \left. + \dis{\int~d^8z~{I_M}^A~{E_A}^M + 
\int~d^8z~{J_M}^A{}_B~{\phi^M}_A{}^B+  
i\int~d^8z~\eta^L c_L
+ i\int~d^8z~\rho^A b_A } \right. \label{path-gravity}\\
& \left. + ~\dis{i\int~d^8z~{K_M}^A{\cal L}_c {E_A}^M +
 i\int~d^8z~{K''_M}^A{}_B{\cal L}_c~{\phi^M}_A{}^B+ 
\int~d^8z~L_M{\cal L}_c{c^M}}\right], 
\no
\end{eqnarray}
where new external sources ${K_M}^A,$  ${K''_M}^A{}_B$ and $L_M$ coupled 
to the BRST variations of the vielbein, the spin connection and the ghost 
under group of diffeomorphisms are introduced. The action 
(\ref{total-action-gravity}) is invariant with respect to BRST transformations
\cite{BRST}, 
\begin{eqnarray}
& \dis{{E_A}^M  \rar {E_A}^M + i{\cal L}_c {E_A}^M\ve },  \no \\
& \dis{{\phi^M}_A{}^B  \rar {\phi^M}_A{}^B + i{\cal L}_c 
{\phi^M}_A{}^B\ve },  \no \\
& \dis{c \rar c - \frac{1}{2}{\cal L}_c{c}\ve}, \no \\
& \dis{b_A \rar b_A + \frac{1}{\a}\le \pd_M {E_A}^M \ri\ve.}  
\label{BRST}
\end{eqnarray}
This symmetry exists due to the property of Lie derivative for any three 
world vectors  
\begin{eqnarray*}
{\cal L}_{\xi}{\cal L_{\psi}}\chi^M + {\cal L}_{\chi}{\cal L_{\xi}}\psi^M
+ {\cal L}_{\psi}{\cal L_{\chi}}\xi^M = 0,  
\end{eqnarray*}
where  
\begin{eqnarray*}
{\cal L}_{\xi}\eta^M = \xi^N(\pd_N\eta^M) - (\pd_N\xi^M)\eta^N.  
\end{eqnarray*}
The effective action $\G$ is related to $W = i~ln~Z$ by the Legendre
transformation
\begin{eqnarray}
& \dis{{E_A}^M \equiv - \frac{\del W}{\del {I_M}^A},~~~
{\phi^M}_A{}^B \equiv - \frac{\del W}{\del {J_M}^A{}_B}, 
  ~~~ ic^L  \equiv - \frac{\del W}{\del \eta_L}, ~~
  ib^A \equiv - \frac{\del W}{\del \rho_A}}, ~~\label{defphi}\\
& \dis{\G = - W - \int~d^8z~{I_M}^A~{E_A}^M 
- \int~d^8z~{J_M}^A{}_B~{\phi^M}_A{}^B 
- i\int~d^8z~\eta^L c_L - i\int~d^8z~\rho^A b_A}\no 
\end{eqnarray}
If all equations Eq.~(\ref{defphi}) can be inverted,
\begin{eqnarray*}
& \Omega = \Omega[\varphi,{K_M}^A,{K''_M}^A{}_B, L_M], \\
& \dis{\Omega \equiv \le {I_M}^A, {J_M}^A{}_B,\eta_L,\rho_A\ri,  ~~~   
\varphi \equiv \le {E_A}^M , {\phi^M}_A{}^B, c^L, b^A\ri}. \no
\end{eqnarray*}
the effective action can be  defined in terms of new variables,
$\G = \G[\varphi,{K_M}^A,{K''_M}^A{}_B,L_M].$
Hence the following equalities hold:
\begin{eqnarray}
& \dis{\frac{\del \G}{\del {E_A}^M} = - {I_M}^A, ~~~ 
 \frac{\del \G}{\del {\phi^M}_A{}^B} =  - {J_M}^A{}_B, ~~~
\frac{\del \G}{\del {K_M}^A} = - \frac{\del W}{\del {K_M}^A}}, 
\label{GW}\\
& \dis{\frac{\del \G}{\del {K''_M}^A{}_B} 
= - \frac{\del W}{\del {K''_M}^A{}_B,}, ~~~
\frac{\del \G}{\del c^L} = i\eta_L, ~~ \frac{\del \G}{\del b^A} = 
i\rho_A,  ~~ \frac{\del \G}{\del L_M} = - \frac{\del W}{\del L_M}}. \no
\end{eqnarray}

If the transformation Eq.~(\ref{BRST}) is made in the path integral 
Eq.~(\ref{path-gravity}) one obtains (as the result of the invariance of the
integral Eq.~(\ref{path-gravity}) under a change of variables) the 
Slavnov--Taylor (ST) identity (up to dependent on $\beta$ terms):
\begin{eqnarray*}
& \dis{\left[\int~d^8z~{I_M}^A\frac{\del}{\del{K_M}^A}
+ \int~d^8z~{J_M}^A{}_B\frac{\del }{\del {K''_M}^A{}_B}
- \int~d^8z~i\eta_M\le\frac{\del}{\del L_M}\ri \right.} \no\\
& + \dis{\left. 
\int~d^8z~i\rho^A\le\frac{1}{\a}\pd_M\frac{\del}{\del {I_M}^A}\ri
\right]W} = 0,
\end{eqnarray*}
or, taking into account the relations Eq.~(\ref{GW}), we have
\begin{eqnarray}
& \dis{\int~d^8z~\frac{\del \G}{\del {E_A}^M}\frac{\del \G}{\del {K_M}^A} + 
\int~d^8z~\frac{\del \G}{\del  {\phi^M}_A{}^B}
\frac{\del \G}{\del {K''_M}^A{}_B}
+ \int~d^8z~\frac{\del \G}{\del c^M}\frac{\del \G}{\del L_M}} \no\\
& \dis{- \int~d^8z~\frac{\del \G}{\del b_A}\le\frac{1}{\a}\pd_M {E_A}^M\ri}
 = 0.
\label{STrM}
\end{eqnarray}
In addition to ST identity also there is the ghost equation
that can be derived by shifting the antighost field $b$ by an arbitrary 
field $\ve(z)$ in the path integral. The consequence of 
invariance of the path integral with respect to such a change of 
variable is (in terms of the variables (\ref{defphi})) \cite{SF}
\begin{eqnarray}
\frac{\del \G}{\del b^A(z)} + \pd_M~\frac{\del \G}{\del {K_M}^A(z)} = 0.
 \label{ghost}
\end{eqnarray}
The ghost equation (\ref{ghost}) restricts 
the dependence of $\G$ on the antighost field $b$ and on the external 
source $K_M$ to an arbitrary dependence on their combination 
\begin{eqnarray}
\pd_M~b^A(z) +  {K_M}^A(z). \label{comb}
\end{eqnarray}
Starting with this point we can use method proposed in Refs. 
\cite{jhep,CKS1,CKS2} for searching solution to Slavnov--Taylor identity 
in theories with local gauge symmetries. The main idea of  Refs. 
\cite{jhep,CKS1,CKS2} is to take a solution in which 1PI $Lcc$ correlator 
of the theory is invariant itself with respect to Slavnov--Taylor identity.
The $Lcc\phi_M$ and $LccE_M$ correlators 
have more weak behaviour in space of momentums and their contribution 
to the ST identity will be more weak in comparison with $Lcc$ contribution.
This results in invariance of the effective action which is local 
construction written in terms of dressed effective fields with respect to the 
BRST symmetry of diffeomorphisms (\ref{BRST}) written also in terms of 
dressed effective fields.

At the same time, there is another symmetry for the structure group. 
It has been analyzed in Refs. \cite{jhep,CKS1,CKS2} for the case of Yang--Mills 
theory and can be repeated here without modifications since all the 
constructions for the spin connection just repeat the analogous construction for 
the Yang--Mills connection. The field of vielbein participates in that 
symmetry as well as another matter field. According to the lines of that 
approach, the effective action in supergravity theory that is in consideration 
is the following:    
\begin{eqnarray}
& \dis{\G[\pi^\omega,{E_M}^A,{\phi_{MA}}^B,b,c,b',c']} = 
\dis{ -\frac{6}{\kappa^2}\int~d^4y~d^2{\Theta}~
{\cal E}({{\tilde{E}}_M}{}^A,{{\tilde{\phi}}_{MA}}{}^B)
R({{\tilde{E}}_M}{}^A,{{\tilde{\phi}}_{MA}}{}^B) + \dots} \no\\
& \dis{- \int~d^8z ~\frac{1}{2\a}\le \pd_M {E_A}^M \ri^2 
-\int~d^8z~\Tr\le\frac{1}{\b}
\left[\pd_M\phi^M(z)\right]^2 \ri - 
\int~d^8z ~i~\tilde{b}^A~\pd_M~{\cal L}_{\tilde{c}}{{\tilde{E}}_A}{}^M} \no\\ 
& \dis{-\int~d^8z~2~\Tr~\le~i~
\tilde{b'}(z)\pd_M~\tilde{\cal D}^M~\tilde{c'}(z)\ri
+ \int~d^8z~\tilde{E}~\pi^\omega\tilde{\cal C}^\omega}, \label{G} 
\end{eqnarray}   
where all auxiliary fields $K$ and $L$ are set equal to zero. As one can see,
the physical part of the effective action is local BRST invariant with 
respect to both local symmetries (Lorentz and diffeomorphism). Physical part 
starts with the first term of (\ref{G}) and contains all other possible 
invariants in terms of chiral density, chiral superfield $R,$ vielbein and 
spin connection. It is unclear at present how to derive exact form of the 
physical part. Dressed fields in the 
effective action are the effective fields convoluted with unspecified dressing 
functions, 
\begin{eqnarray*} 
& \dis{ {\tilde{E}}_M{}^A(z)  = \int~d~z'~G^{-1}_{E}(z-z')~{E_M}^A(z'),} \\   
& \dis{ {\tilde{\phi}_{MA}}{}^B(z)  = 
  \int~d~z'~G^{-1}_{\phi}(z-z')~{\phi_{MA}}^B(z'),} \\ 
& \dis{ \tilde{c}(z)   = \int~d~z'~G^{-1}_{c}(z-z')~c(z') }, \\
& \dis{ \tilde{b}(z) = \int~d~z'~G_{E}(z-z')~b(z') }.
\end{eqnarray*} 
     
In the effective action (\ref{G}) we have not done the integration yet over the 
Lagrange multiplier $\pi.$ In 1PI diagrams this factor can be considered 
as background superfield. By requiring correspondence to the classical 
action we obtain that the constraint  (\ref{constraint}) has been modified to
the following form:    
\begin{eqnarray*}
\pi^{\omega}\le T - \tilde{\cal{D}}\tilde{E}\ri^{\omega}.      
\end{eqnarray*}
Integration over $\pi$ in the path integral means that we have to resolve 
this constraint in the effective action. In comparison with the classical 
action we can derive the analogous solution with the only difference that
instead of classical fields of the vielbein and spin connection we have to 
resolve it for the dressed vielbein and the dressed spin connection. 

We comment here that this action should be considered only as one of the models
for quantum supergravity. We do not pretend in this note for strict 
argument in favor of this action. However, this idea to write the effective 
action in 
terms of {\em dressed} (convoluted with some unspecified dressing 
functions) effective fields seems plausible physically and probably has 
natural physical interpretation.

\vskip 3mm 
\noindent {\large{\bf{Acknowledgments}}} 
\vskip 3mm

Financial support of Mineduc (Chile) under grant FSM9901 and Conicyt (Chile) 
under grant 8000017 and 1040368 is acknowledged. I am grateful to 
Department of Physics of Brown University where part of this work has 
been done for the kind hospitality and financial support during my stay.

\end{document}